\documentstyle[11pt,newpasp,twoside,epsf]{article}
\newcommand\Mwd   {M_{\rm wd}}
\newcommand\Rwd   {R_{\rm wd}}
\newcommand\Msun  {{\rm M_{\odot }}}
\newcommand\Mdot  {\dot{M}}

\def\edcomment#1{\iffalse\marginpar{\raggedright\sl#1\/}\else\relax\fi}
\reversemarginpar

\begin{document}

\title{EUVE Observations of Nonmagnetic Cataclysmic
       Variables\altaffilmark{1}}
\author{Christopher W.~Mauche}
\affil{Lawrence Livermore National Laboratory,
       L-43, 7000 East Avenue, Livermore, CA 94550}

\altaffiltext{1}{Dedicated to the memory and accomplishments of Danie
Overbeek, a charming man and devoted ``amateur'' observer, who, over
a period of nearly 50 years, contributed over 287,000 observations of
variable stars to the AAVSO International Database.}

\setcounter{footnote}{1} 

\begin{abstract}
We summarize {\it EUVE\/}'s contribution to the study of the boundary
layer emission of high accretion-rate nonmagnetic cataclysmic variables,
especially the dwarf novae SS~Cyg, U~Gem, VW~Hyi, and OY~Car in outburst.
We discuss the optical and EUV light curves of dwarf nova outbursts, the
quasi-coherent oscillations of the EUV flux of SS~Cyg, the EUV spectra of
dwarf novae, and the future of EUV observations of cataclysmic variables.
\end{abstract}

\section{Introduction}

Cataclysmic variables (CVs) are a diverse class of semidetached binaries
composed of a low-mass main-sequence star and an accreting white dwarf. In
nonmagnetic CVs conservation of angular momentum dictates that accretion
onto the white dwarf is mediated by a disk. While the white dwarf
and disk are the dominant sources of optical through FUV light, the
boundary layer between the disk and the surface of the white dwarf is the
dominant source of higher-energy emission. Simple theory predicts that
the accretion disk and boundary layer luminosities should be comparable,
with $L_{\rm disk}\approx L_{\rm bl}\approx G\Mwd\Mdot/ 2\Rwd\sim 3\times
10^{34}\, (\Mdot/10^{-8}~\rm \Msun~yr^{-1})~erg~s^{-1}$, where $\Mdot $
is the accretion rate and $\Mwd $ and $\Rwd $ are respectively the mass
and radius of the white dwarf. When $\Mdot $ is low ($\Mdot\sim
10^{-11}~\rm \Msun~yr^{-1}$, as in dwarf novae in quiescence), the
boundary layer is optically thin and quite hot (of order the virial
temperature $T_{\rm vir}=G\Mwd m_{\rm H}/3k\Rwd\sim 10$ keV); when
$\Mdot $ is high ($\Mdot\sim 10^{-8}~\rm \Msun~yr^{-1}$, as in novalike
variables and dwarf novae in outburst), the boundary layer is optically
thick and quite cool (of order the blackbody temperature $T_{\rm bb}=
[G\Mwd\Mdot/8\pi\sigma\Rwd ^3]^{1/4}\sim 10$ eV). Hence, the boundary
layer emission of high-$\Mdot $ CVs is radiated primarily in the EUV,
where it is easily hidden from us by the interstellar medium.

In addition to the severe effect of photoelectric absorption in the EUV,
progress in our understanding of the EUV/soft X-ray emission of
high-$\Mdot $ CVs has been hampered by the poor energy resolution of
X-ray detectors, the target-of-opportunity (TOO) scheduling required to
observe dwarf novae in outburst, and the long (10--30 day) exposures
required to follow the evolution of dwarf novae outbursts. Absorption
by the interstellar medium restricts us to a relatively small number
of intrinsically bright, nearby systems with low interstellar column
densities ($N_{\rm H}\la 10^{20}~\rm cm^{-2}$). The {\it EUVE\/}
spectrometers provide a significant improvement in the spectral resolution
of proportional counter and microchannel plate detectors of past missions,
and although the {\it EUVE\/} spectrometer effective areas were small, the
necessarily long integration times made it possible for the first time to
follow the evolution of all or part of several dwarf nova outbursts. {\it
EUVE\/}'s ability to observe dwarf novae in outburst was made possible on
one hand by the members, staff, and directors of the American Association
of Variable Star Observers (AAVSO) and the Variable Star Section/Royal
Astronomical Society of New Zealand (VSS/RASNZ), who initiated the TOO
requests, and the staff of the {\it EUVE\/} Science Operations Center at
CEA and the Flight Operations Team at GSFC, who quickly reacted to them.

{\it EUVE\/} was used during its lifetime to observe among other
nonmagnetic CVs the dwarf novae SS~Cyg in narrow and wide, normal and
anomalous outbursts; U~Gem in normal outburst (twice); VW~Hyi in normal
and super\-outburst (twice); OY~Car in super\-outburst (twice), T~Leo in
super\-outburst (Howell et al.\ 1999), and the novalike variable IX~Vel
(van Teeseling et al.\ 1995). These observations were obtained for
varying reasons, and on a couple of occasions they were coordinated with
other satellites ({\it HST\/}, {\it Voyager\/}, {\it Chandra\/}, {\it
RXTE\/}) sensitive in other wavebands (UV, FUV, X-rays). Details of the
{\it EUVE\/} observations are listed in Table~1.

\begin{table}
\caption{Journal of {\it EUVE\/} Observations of Nonmagnetic CVs}
\begin{tabular}{lcccll}
\tableline
    &  Date&	      Interval&	Exp.&	Type of    &        \\
Star&	(M/Y)&	(JD$-$2400000)&	(ks)&	Outburst&	Comment\\
\tableline
SS Cyg&	08/93&	49216.58--222.86&	          179.4&  Anom.~Wide&    \\
IX Vel& 11/93& 49317.99--326.85&           222.4&  \ldots &       \\
 U Gem&	12/93&	49350.00--361.15&	          249.0&  Normal&        \\
VW Hyi&	06/94&	49505.46--507.66&	\phantom{0}89.4&  Super&         \\
SS Cyg&	06/94&	49526.67--536.69&	          147.8&  Normal Wide&   \\
VW Hyi&	07/95&	49906.70--917.29&	          183.8&  Normal&        + {\it Voyager\/}\\
VW Hyi&	05/96&	50210.58--218.47&	\phantom{0}55.4&  Super&         + {\it RXTE\/}   \\
SS Cyg&	10/96&	50366.40--379.45&	          208.1&  Normal Narrow& + {\it RXTE\/}   \\
 T Leo&	02/97&	50500.60--506.54&	\phantom{0}96.1&  Super&         + {\it RXTE\/}   \\
OY Car&	03/97&	50534.46--537.64&	\phantom{0}94.8&  Super&         \\
 U Gem&	11/97&	50760.27--766.85&	          150.0&  Normal&        + {\it RXTE\/}   \\
SS Cyg&	06/99&	51336.84--349.67&	          274.0&  Anom.~Narrow&  + {\it RXTE\/}\\
OY Car&	02/00&	51597.66--601.26&	\phantom{0}69.1&  Super&         \phantom{+ }{\small \& }{\it HST\/}\\
SS Cyg& 09/00& 51800.19--801.96& \phantom{0}42.8&  Normal&        + {\it Chandra\/}\\
\tableline
\end{tabular}
\end{table}

We provide highlights of various aspects of these observations in the next
sections, limiting attention to sources which were bright enough to yield
detailed EUV light curves and spectra: specifically, the dwarf novae
SS~Cyg, U~Gem, VW~Hyi, and OY~Car in outburst. The subsequent sections
discuss the optical and EUV light curves of dwarf novae (\S 2), the
quasi-coherent oscillations of the EUV flux of SS~Cyg (\S 3), and the EUV
spectra of dwarf novae (\S 4). We close in \S 5 with comments about the
future of EUV observations of CVs.

\section{Optical and EUV Light Curves}

Optical light curves of outbursts of SS~Cyg, U~Gem, VW~Hyi, and OY~Car
were constructed from visual magnitude estimates and CCD photometric
measurements obtained by members of the AAVSO and VSS/RASNZ. EUV light
curves were constructed from {\it EUVE\/} deep survey photometer (DS)
data or short wavelength spectrometer (SW) data in those instances when
the DS was turned off (during the peak of the 1996 October outburst of
SS~Cyg, both outbursts of U~Gem, and the 1994 June outburst of VW~Hyi).
The full set of optical and EUV light curves is shown in Mauche, Mattei,
\& Bateson (2001); in Figures 1--3 we reproduce the light curves of
normal outbursts of SS~Cyg, U~Gem, and VW~Hyi. In Figure~3 the {\it
Voyager\/} FUV (950--1150~\AA ) flux density light curve of VW~Hyi is
shown by the filled triangles. In the lower-left panel of Figure~1 the
{\it RXTE\/} hard X-ray (2--10 keV) count rate light curve of SS~Cyg
(Wheatley, Mauche, \& Mattei 2000) is shown by the open triangles.

These light curves provide important diagnostics of the nature of dwarf
nova outbursts because the optical, UV, and EUV flux is produced in
physically distinct regions: the optical flux in the outer disk, the UV
flux in the inner disk, and the EUV flux in the boundary layer. Dwarf
nova outbursts are due to an instability in the rate of mass transfer
through the disk caused by the dramatic change in opacity when H becomes
partially ionized at $T\sim 10^4$ K (for reviews of the disk instability
model, see Cannizzo 1993; Osaki 1996; Lasota 2001). The instability can
be triggered at large or small disk radii, resulting in normal, fast-rise
outbursts or anomalous, slow-rise outbursts, respectively. In either
case, the beginning of the outburst is signaled by a rise of the optical
flux, followed by a rise of the UV flux as material sinks through the
disk, converting its gravitational potential energy into rotational
kinetic energy and radiation. This is followed by a rise in the EUV flux
as material passes through the boundary layer, where its prodigious
rotational kinetic energy is converted into radiation.

The anomalous outbursts of SS~Cyg (upper panels of Fig.~1) manifest the
gradual increase of the optical and EUV light curves expected for
outbursts triggered near the inner edge of the disk. The optical and EUV
fluxes rise during the beginning of these outbursts as the heating wave
which transforms the disk from quiescence to outburst sweeps outward,
causing more and more material to flow through the disk and boundary
layer onto the white dwarf. In contrast, the normal outbursts of SS~Cyg
(lower panels of Fig.~1), U~Gem (Fig.~2), and VW~Hyi (Fig.~3) manifest
the fast increase of, and the delay between, the optical and EUV light
curves expected for outbursts triggered near the outer edge of the disk.
As measured from the initial rise of the optical light curves, the delay
of the rise of the EUV light curves is $\approx 1.5$, 1.25, and 0.75 days
for SS~Cyg, U~Gem, and VW~Hyi, respectively. In VW~Hyi, the FUV light
curve rises $\approx 0.5$ days after the optical light curve and $\approx
0.25$ days before the EUV light curve, but falls as slowly as the optical
light curve, consistent with the expectation that the accretion disk and
not the boundary layer is the source of the FUV flux.

\begin{figure}
\plotone{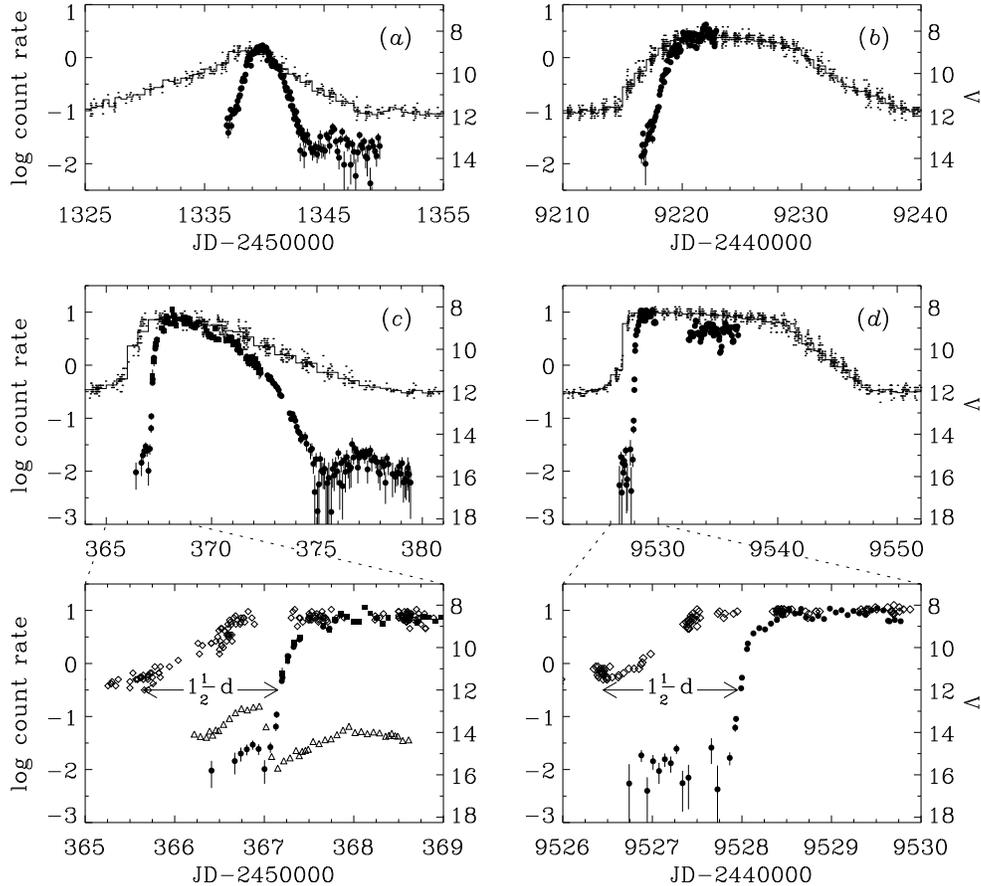}
\caption{Optical, {\it EUVE\/}, and {\it RXTE\/} light curves of the
anomalous ({\it a\/}) 1999 June and ({\it b\/}) 1993 August outbursts and
the normal ({\it c\/}) 1996 October and ({\it d\/}) 1994 June outbursts
of SS~Cyg. DS and scaled SW count rates are shown by the filled circles
and squares, respectively; {\it RXTE\/} PCA count rates (scaled downward
by 2.5 dex) are shown by the open triangles; individual AAVSO
measurements are shown by the small dots and diamonds; half-day mean
optical light curve is shown by the histogram. For the normal outbursts
of SS~Cyg the optical-EUV delay is $\approx 1.5$ days.}
\end{figure}

\begin{figure}
\plotone{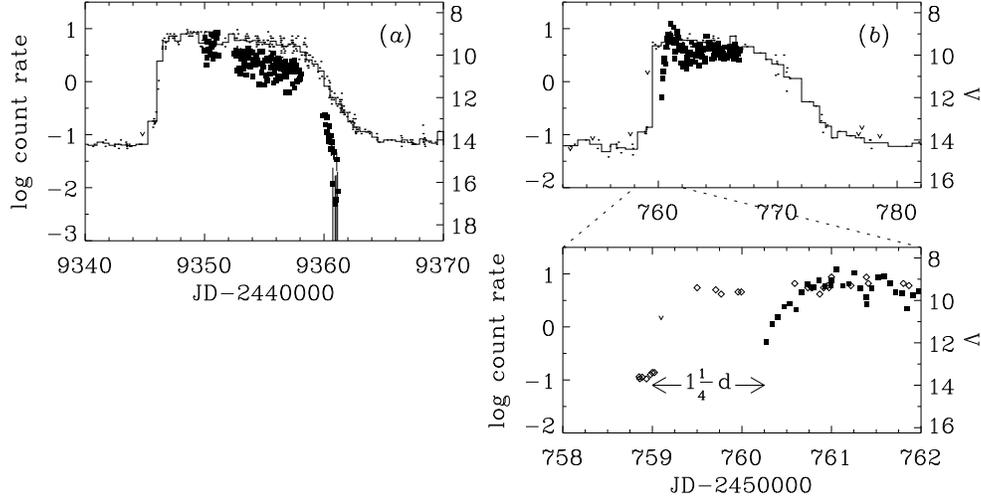}
\caption{Optical and {\it EUVE\/} light curves of the normal ({\it a\/})
1993 December and ({\it b\/}) 1997 November outbursts of U~Gem. SW count
rates are shown by the filled squares, individual AAVSO measurements are
shown by the small dots and diamonds, half-day mean optical light curve
is shown by the histogram. The optical-EUV delay of U~Gem is $\approx
1.25$ days.}
\end{figure}

\begin{figure}
\plotone{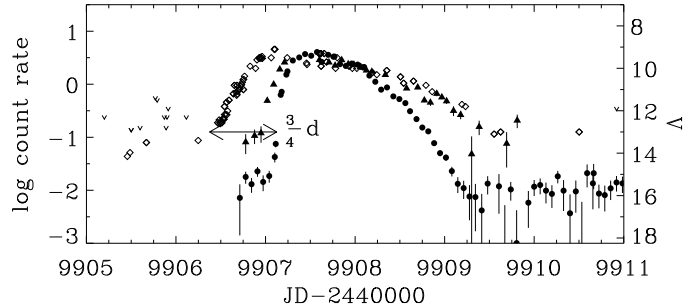}
\caption{Optical, {\it Voyager\/}, and {\it EUVE\/} light curves of the
normal 1995 July outburst of VW~Hyi. DS count rates are shown by the
filled circles, {\it Voyager\/} FUV flux densities are shown by the filled
triangles, individual VSS/RASNZ and AAVSO measurements are shown by the
diamonds. The optical-FUV delay of VW~Hyi is $\approx 0.5$ days and the
optical-EUV delay is $\approx 0.75$ days.}
\end{figure}

As detailed by Mauche, Mattei, \& Bateson (2001), the optical-EUV delays
of dwarf novae can be used to estimate the velocity of the heating
wave, and hence the value of the viscosity parameter of high-$\Mdot $
CVs.\footnote{Upsetting this simple picture is the early rise of the
hard X-ray light curve of the 1996 October outburst of SS~Cyg (lower-left
panel of Fig.~1). This signals that {\it some\/} increase in the
mass-accretion rate through the boundary layer occurs during the early
stages of the outburst, but the increase is not large: the hard X-ray
flux rose by a factor of only $\sim 4$, whereas the EUV flux rose by a
factor of $\sim 10^3$ when the boundary layer made the transition from
optically thin to thick.} Assuming the system parameters (orbital period,
white dwarf mass, and mass ratio) tabulated by Ritter \& Kolb (1998),
that the radius of the disk $R_{\rm disk}\approx 0.7\times R_{\rm L1}$,
and that the disk instability starts at the outer edge of the disk, the
velocity of the heating wave $v\approx R_{\rm disk} /{\rm delay}\approx
3~\rm km~s^{-1}$. This result is consistent with $v=\alpha c_{\rm s}$ if
the viscosity parameter $\alpha\approx 0.2$ and the sound speed $c_{\rm
s} =10\, (T/10^4~{\rm K})^{1/2}~\rm km~s^{-1}\approx 15~\rm km~s^{-1}$.
For a more in-depth discussion of this topic, see Cannizzo (2001).

\section{Quasi-Coherent Oscillations of SS~Cyg}

Another use of the {\it EUVE\/} photometric measurements of dwarf novae
is studies of their high-frequency temporal variability: the
quasi-periodic and quasi-coherent oscillations known from
previous EUV/soft X-ray observations. As in previous {\it HEAO~1\/} LED~1
(C\'ordova et al.\ 1984) and {\it EXOSAT\/} LE (Mason et al.\ 1988) data,
$\sim 25$ s low-amplitude low-coherence quasi-periodic oscillations were
observed in {\it EUVE\/} SW photometry of U~Gem in outburst (Long et al.\
1996). van der Woerd et al.\ (1987) detected $\sim 14$ s quasi-coherent
oscillations on two occasions in {\it EXOSAT\/} LE data of VW~Hyi in
superoutburst, but it proved impossible to detect oscillations in the
{\it EUVE\/} data, even though the source was observed extensively.

\begin{figure}
\plotone{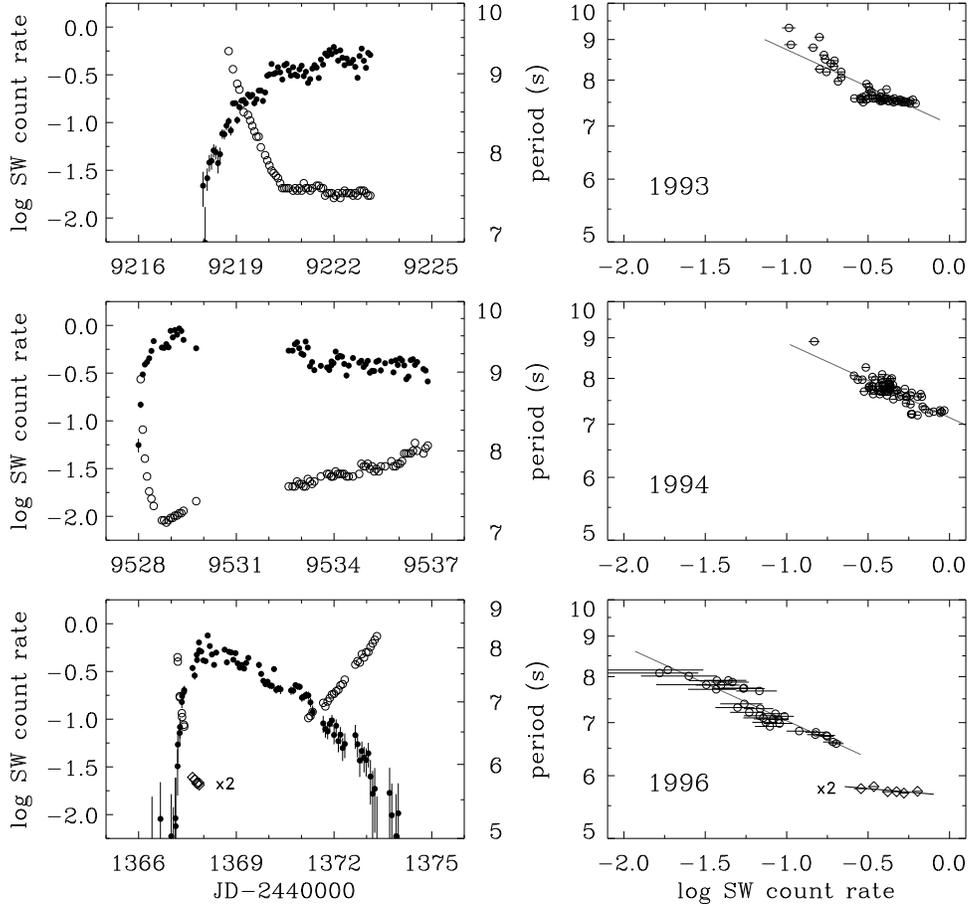}
\caption{{\it Left panels\/}: SW count rate ({\it filled circles with
error bars\/}) and period of the EUV oscillation ({\it open symbols\/})
as a function of time for the 1993 August, 1994 June, and 1996 October
outbursts of SS~Cyg. {\it Right panels\/}: Period $P$ as a function of
SW count rate $I$; straight gray lines are $P\propto I^{-0.094}$~s and
$P\propto I^{-0.018}$. Open diamonds in the lower panels are plotted at
twice the observed periods.}
\end{figure}

Far more rewarding is SS~Cyg, whose EUV/soft X-ray quasi-coherent
oscillations have been detected in {\it HEAO~1\/} LED~1 (C\'ordova et
al.\ 1980, 1984), {\it EXOSAT\/} LE (Jones \& Watson 1992), and {\it
ROSAT\/} HRI (van Teeseling 1997) data. Various aspects of the rich
phenomenology of the EUV oscillations of SS~Cyg are discussed by Mauche
(1996a, 1997a, 1997b, 1998) and Mauche \& Robinson (2001). We show in
Figure~4 a compilation of the variations of oscillation period versus time
and oscillation period versus EUV flux for the 1993 August, 1994 June, and
1996 October outbursts of SS~Cyg. A tight correlation between oscillation
period $P$ and SW count rate $I$ is apparent in the right panels of the
figure, with $P\propto I^{-0.094}$. Two conclusions are immediately
apparent from these figures. First, we see that the period of the EUV
oscillations of SS~Cyg is a single-valued function of the EUV flux
(hence, by inference, the mass-accretion rate onto the white dwarf). The
loops observed in plots of oscillation period versus optical flux (e.g.,
Patterson 1981) are understood to be the result of the delay between
the rise of the optical and EUV flux at the beginning of dwarf nova
outbursts (\S 2). Second, the period-intensity variation of SS~Cyg is
far ``stiffer'' than that expected for disk accretion onto a compact
star with a dipole magnetic field, for which $P\propto\Mdot ^{-3/7}$.
If such a model applies to SS~Cyg, an effective high-order multipole
magnetic field is required, with a field strength $B\sim 0.1$--1 MG at
the surface of the white dwarf (Mauche 1996a).

An additional phenomenon illustrated in the lower-left panel of Figure~4
is the frequency doubling observed during the 1996 October outburst of
SS~Cyg: on the rise to outburst the period of the EUV oscillation was
observed to {\it jump\/} from 6.59~s to 2.91~s (Mauche 1998). Frequency
doubling has never been observed before in this or any other dwarf
nova in either the optical or EUV/soft X-rays, although similarly
short-period ($P=2.8$--2.9~s) oscillations were detected by van Teeseling
(1997) in {\it ROSAT\/} HRI data acquired during the 1996 December
outburst of SS~Cyg. As the lower-right panel of Figure~4 makes clear,
after the oscillation frequency doubled, its dependence on the SW count
rate became ``stiffer'' by a factor of $\approx 5$ in the exponent
($P\propto I^{-0.018}$). SS~Cyg seems to have been doing what it could
to avoid oscillating faster than about 2.8~s. If this is the Keplerian
period of material at the inner edge of the accretion disk, then $P_{\rm
Kep}\ge 2\pi (\Rwd ^3/G\Mwd)^{1/2}\approx 2.8$~s, requiring $\Mwd\ga
1.27~\Msun $. If instead, $P_{\rm Kep}\approx 5.6$~s (i.e., the observed
2.8~s period is the first harmonic of a 5.6~s Keplerian period), then
$\Mwd\ga 1.08~\Msun $. Existing radial velocity data favor the second
option, but it requires only a $\approx 10\%$ reduction in the inclination
angle to accommodate the first option. A secure white dwarf mass is
required to determine whether the 2.8~s oscillation is the fundamental or
the first harmonic of the intrinsic oscillation.

During the same observations, Mauche \& Robinson (2001) observed for the
first time in SS~Cyg or any other dwarf nova quasi-coherent oscillations
{\it simultaneously\/} in the optical and EUV. They found that the period
and phase of the oscillations are the same in the two wavebands, finally
confirming the long-held assumption that the periods of the optical and
EUV/soft X-ray oscillations of dwarf novae are equal. The $UBV$
oscillations can be simply the Rayleigh-Jeans tail of the EUV oscillations
if during outburst the boundary layer temperature $kT\la 15$~eV and hence
the luminosity $L_{\rm bl}\ga 1.2\times 10^{34}\, (d/{\rm 75~pc})^2~\rm
erg~s^{-1}$ (comparable to that of the accretion disk). Otherwise, the
lack of a phase delay ($\Delta\phi _0 =0.014\pm 0.038$ or $\Delta
t=0.10\pm 0.26$~s) between the EUV and optical oscillations requires that
the optical reprocessing site lies within the inner third of the accretion
disk. This is strikingly different from other CVs, where much or all of
the disk contributes to the optical oscillations.

\begin{figure}
\plotone{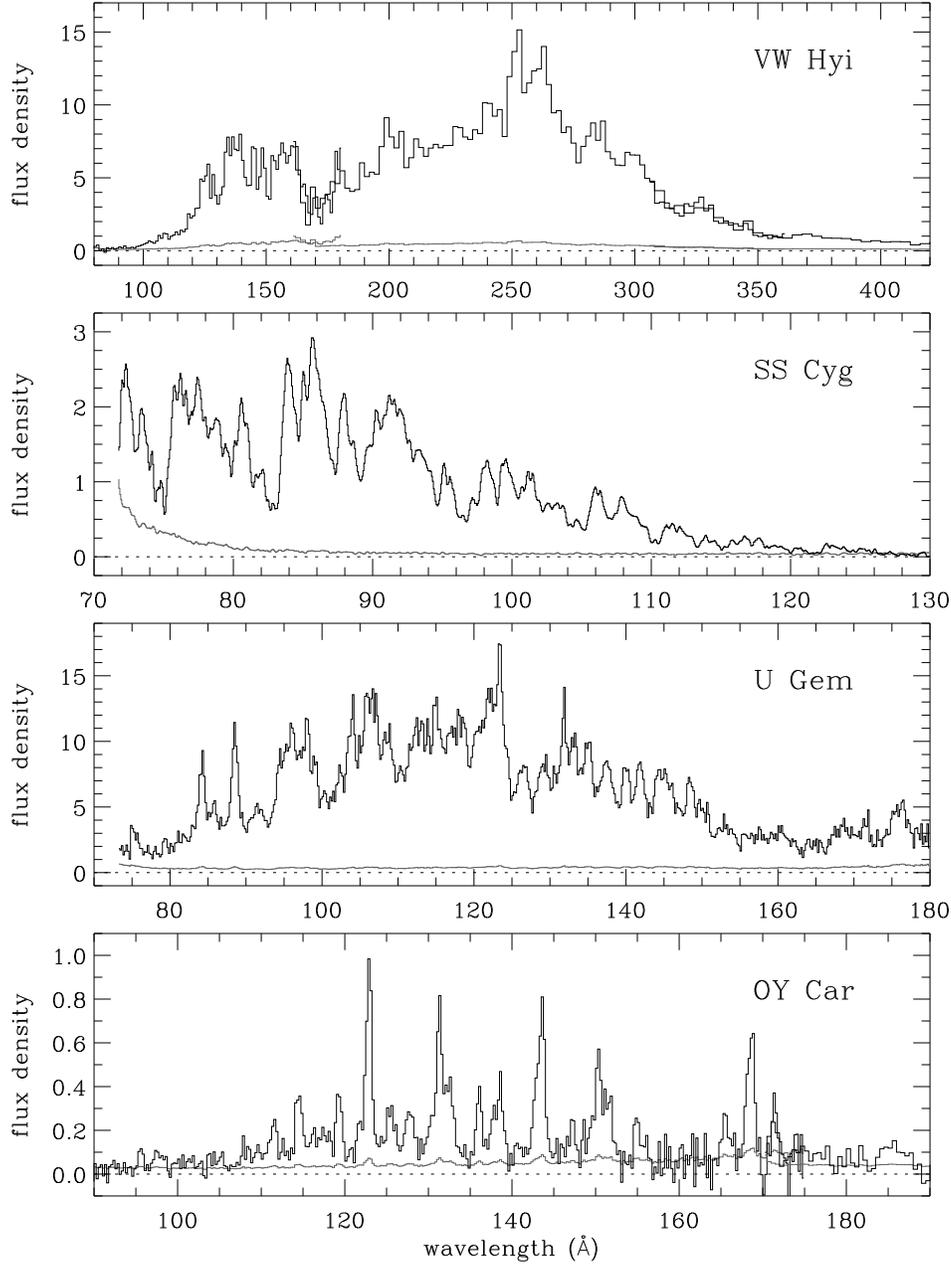}
\caption{{\it EUVE\/} spectra of VW~Hyi, SS~Cyg, U~Gem, and OY~Car.
Lower gray histograms are the $1\, \sigma $ error vectors. Units of
flux density are $10^{-12}~\rm erg~cm^{-2}~s^{-1}.$}
\end{figure}

\section{EUV Spectra}

{\it EUVE\/}'s truly revolutionary advancement over previous EUV/soft
X-ray missions was the resolving power of its grating spectrometers, which
was $\lambda/\Delta\lambda\approx 200$ at 100~\AA , compared to $E/\Delta
E\sim 1$ at 0.2 keV for proportional counter detectors. {\it EUVE\/}'s
resolving power is fairly well matched to CVs, which have Keplerian
Doppler widths $v\la (G\Mwd/\Rwd )^{1/2}\approx 3000~\rm km~s^{-1}$ hence
$\lambda/\Delta\lambda=c/v\ga 100$. From the [$N_{\rm H}$, $kT$] banana
plots of past missions we have moved to the frightening level of detail
shown in Figure~5, which shows representative {\it EUVE\/} spectra of
SS~Cyg (Mauche, Raymond, \& Mattei 1995), U~Gem (Long et al.\ 1996),
VW~Hyi (Mauche 1996b), and OY~Car (Mauche \& Raymond 2000) in outburst.
The spectral range runs from $\lambda\la 350$~\AA \ for VW~Hyi with
$N_{\rm H}\sim 10^{18}~\rm cm^{-2}$ to $\lambda\la 130$~\AA \ for SS~Cyg
with $N_{\rm H}\sim 4\times 10^{19}~\rm cm^{-2}$, and with the exception
of SS~Cyg none of the sources have any significant flux shortward of
$\lambda\approx 80$~\AA . This, combined with the severe effect of
photoelectric absorption by the interstellar medium, is the explanation
for the ``nova problem,'' {\it aka\/} the ``mystery of the missing
boundary layer,'' as we were assured so long ago by Patterson \& Raymond
(1985).

It is quite clear from Figure~5 that the EUV spectra of dwarf novae in
outburst are not the ``simple'' emission-line spectra of late-type stars
or the continuum spectra of hot white dwarfs. Worse, these four systems
differ markedly from each other, so it is difficult to confidently draw
general conclusions. No attempt has been made to quantitatively understand
the quasi-continuum spectrum of VW~Hyi. The EUV spectrum of SS~Cyg is only
poorly understood. The broad-band spectral energy distribution can be
parameterized by a $kT=20$--30 eV blackbody absorbed by a column $N_{\rm
H}=7.0$--$4.4\times 10^{19}~\rm cm^{-2}$ (in excess of the interstellar
value), but the inferred luminosities are distressingly low: $L_{\rm bl}
=20$--$5\times 10^{32}\, (d/{\rm 75~pc})^2~\rm erg~s^{-1}$, whereas the
disk luminosity $L_{\rm disk}\approx 3\times 10^{34}\, (d/{\rm
75~pc})^2~\rm erg~s^{-1}$. Tentative identifications can be made of some
of the apparent emission features (Mg~VII $2p^2$--$2p3d$ at 83.5--84.0
\AA , Si~VII $2p^4$--$2p^33s$ at 85.3--85.7~\AA , Ne~VIII $2s$--$3p$ at
88.1~\AA \ and $2p$--$3d$ at 98.2~\AA ), but mostly we have failed in this 
effort. Some of the spectral features look like the P~Cygni profiles seen
in UV spectra of SS~Cyg and other high-$\Mdot $ CVs, suggesting that a
significant modification of the intrinsic boundary layer spectrum occurs
as the photons propagate through the system's accretion disk wind. It may
be the case, as in novae, that the apparent emission features in the EUV
spectrum of SS~Cyg are simply regions of relative transparency is a sea
of overlapping emission lines. This impression is supported by the {\it
Chandra\/} LETG (1--170~\AA ) spectrum obtained during the 2001 January
outburst of SS~Cyg, but detailed modeling is required to determine if such
a model can be made to work quantitatively.

Compared to the EUV spectrum of SS~Cyg, the EUV spectrum of U~Gem is
relatively simple: it appears to consist of a series of moderately broad
($\rm FWHM\approx 1$--2~\AA ) emission lines superposed on a moderately
strong continuum parameterized by a $kT\approx 11$~eV blackbody absorbed
by a column $N_{\rm H}\approx 3\times 10^{19}~\rm cm^{-2}$ (close to the
interstellar value). Happily, the inferred luminosity $L_{\rm bl}\approx 3
\times 10^{32}\, (d/{\rm 90~pc})^2~\rm erg~s^{-1}$ is comparable to that
of the disk. Line identifications include nearly all the high oscillator
strength transitions arising from the ground levels of O~VI, Ne~VI--VIII,
Mg~VI--VII, Fe~VII--X, and Fe XXIII. Long et al.\ (1996) do not provide a
quantitative model, but they argue that the lines are formed by scattering
of boundary layer radiation in U~Gem's accretion disk wind. That the
line-forming region is more extended than the continuum is indicated by
the binary-phased EUV spectrum of U~Gem (Mauche 1997c), which shows that
many of the emission lines (particularly Ne~VIII $2s$--$3p$ at 88.1~\AA )
persist through the partial eclipses centered at binary phases $\phi\sim
0.1$ and $\phi \sim 0.65$.

Even more simple is the EUV spectrum of OY~Car in superoutburst, which
consists of moderately broad ($\rm FWHM\approx 1$~\AA ) emission lines of 
N~V, O~V--VI, Ne~V--VII, Mg~IV--VI, Fe~VI--VIII, and Fe XXIII; a slightly
cooler set of ions than that observed in U~Gem. OY~Car is seen edge-on, so
in outburst the white dwarf and boundary layer are hidden from view by the
accretion disk at all orbital phases. Neither {\it EXOSAT\/} (Naylor et
al.\ 1988) nor {\it EUVE\/} saw an eclipse of the EUV flux when the white
dwarf and accretion disk were eclipsed by the secondary, demonstrating
that what EUV flux we observe comes from an extended emission region.
A consistent interpretation of the {\it EUVE\/} photometric and
spectroscopic data is supplied by a model wherein radiation from the
accretion disk and boundary layer is scattered into the line of sight by
the system's accretion disk wind (Mauche \& Raymond 2000). It is possible
to trade off continuum luminosity for wind optical depth, but reasonable
models have a boundary layer temperature $kT\approx 8$--11~eV and a
boundary layer and accretion disk luminosity $L_{\rm bl}= L_{\rm disk}
\la 4\times 10^{34}~(d/{\rm 85~pc})^2~\rm erg~s^{-1}$, corresponding to
a mass-accretion rate $\Mdot\la 10^{-8}~\rm \Msun~yr^{-1}$; an absorbing
column density $N_{\rm H}\approx 1.6$--$3.5\times 10^{19}~\rm cm^{-2}$;
and a wind mass-loss rate $\Mdot _{\rm wind}\la 10^{-10}~{\rm
\Msun~yr^{-1}} \approx 0.01\, \Mdot $. Because radiation pressure alone
falls an order of magnitude short of driving such a wind, magnetic forces
must also play a role in driving the wind of OY~Car in superoutburst.

\section{The Future}

The near future of EUV observations of CVs is bright in one way but dim
in many other ways. On one hand we currently have access to the {\it
Chandra\/} LETG spectrograph, which provides superior spectral resolution
($\Delta\lambda=0.05$~\AA , compared to $\Delta\lambda= 0.5$~\AA \ for
{\it EUVE\/}), modestly higher effective area ($A_{\rm eff}\approx 9~\rm
cm^2$ at 100~\AA , compared to $A_{\rm eff}\approx 2~\rm cm^2$ for {\it
EUVE\/}), and a harder bandpass ($\lambda=1$--170~\AA , which is better
suited to most CVs). {\it Chandra\/} also is in a deep orbit, so avoids
the frequent Earth occultations which bedevil low-Earth-orbit satellites
(this was particularly bad for OY~Car, whose orbital period of 91 min
was annoyingly close to {\it EUVE\/}'s orbital period of 95 min). On
the other hand, it is difficult to get {\it Chandra\/} observing time,
especially for TOO observations, especially for LETG observations, which
are necessarily long because of the modest LETG effective area and the
high background in the HRC-S detector. No future NASA mission is currently
being planned to build on the wide variety of research areas opened up by
{\it EUVE\/}. The proposed {\it KRONOS\/} MIDEX mission, with its
coaligned optical, UV, and X-ray telescopes; long exposures; and rapid
response to targets of opportunity offers some hope for long-term
multiwavelength studies of CVs, but a planned EUV photometric telescope
was not included in the baseline mission for cost reasons. In our rush to
study sources at $z\sim 10$ we risk losing the ability to study beyond
the visual bandpass sources in our own neighborhood.

\acknowledgements

This work would not have been possible without the assistance of the many
individuals who made possible our TOO observations of dwarf novae: the
members and staff of the AAVSO and VSS/RASNZ, AAVSO director J.\ Mattei,
VSS/RASNZ director F.\ Bateson, the staff of the {\it EUVE\/} Science
Operations Center at CEA (particularly Science Planners M.\ Samuel, G.\
Wong, D.\ Meriweather, B.\ Roberts, J.\ McDonald, and M.\ Eckert), and the
Flight Operations Team at GSFC. We thank P.~Wheatley for the {\it RXTE\/}
light curve of SS~Cyg shown in Fig.~1, J.\ Holberg and J.\ Collins
for the {\it Voyager\/} light curve of VW~Hyi shown in Fig.~3, and
W.~Liller for the optical CCD photometry of the rise of VW~Hyi shown
in Fig.~3. This work was performed under the auspices of the U.S.\
Department of Energy by University of California Lawrence Livermore
National Laboratory under contract No. W-7405-Eng-48.

\end{document}